\documentclass{WileyMSP-template}

\usepackage[mathlines]{lineno}

\usepackage{graphicx} 
\usepackage{dcolumn} 
\usepackage{bm,amsmath,amssymb} 
\usepackage{color} 
\usepackage[bookmarks=false]{hyperref}
\hypersetup{colorlinks=true,citecolor=blue,linkcolor=red,%
	urlcolor=blue,pdfstartview=FitH,bookmarksopen=true}
\makeatother
\usepackage{csquotes}
\MakeOuterQuote{"}
\usepackage{soul} 
\soulregister\cite7
\soulregister\ref7 
\soulregister\qzc7 
\usepackage{ragged2e}

\newcommand{\rev}{\textcolor{black}}

\newcommand{\qzc}[1]{\textbf{\textcolor{cyan}{[#1]}}}

\let\oldequation\equation
\let\oldendequation\endequation
\renewenvironment{equation}
{\linenomathNonumbers\oldequation}
{\oldendequation\endlinenomath}

\begin{document}

\pagestyle{fancy}
\rhead{\includegraphics[width=2.5cm]{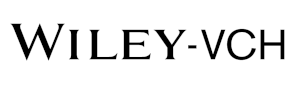}}

\title{Bright Heralded Single-Photon Source Saturating Theoretical Single-photon Purity}

\maketitle


	\author{Haoyang Wang,} 
	\author{Huihong Yuan,} 
	\author{Qiang Zeng*,}
    \author{Lai~Zhou,} 
	\author{Haiqiang Ma,} 
	\author{Zhiliang Yuan}	  
	
	\date{\today}

\dedication{
}

\begin{affiliations}
	
Mr. Haoyang Wang, Dr. Huihong Yuan, Dr. Qiang Zeng, Dr. Lai Zhou, Dr. Zhiliang Yuan\\
Beijing Academy of Quantum Information Sciences, Beijing 100193, China\\
Email Address: \href{zengqiang@baqis.ac.cn}{zengqiang@baqis.ac.cn}

\bigskip
Mr. Haoyang Wang, Prof. Haiqiang Ma\\
School of Science and State Key Laboratory of Information Photonics and Optical Communications, Beijing University of Posts and Telecommunications, Beijing 100876, China\\

\end{affiliations}


\keywords{silicon waveguide, heralded single-photon source, auto-correlation function}

\justifying
\begin{abstract}

    Single-photon source is the cornerstone for modern quantum information processing.
    The present work derives the theoretical limit of single-photon purity for general parametric heralded single-photon sources, \rev{and subsequently} demonstrates a bright, gigahertz-pulsed heralded source with the purity saturating the limit.
    By stimulating spontaneous four-wave mixing effect in the silicon spiral waveguide, this on-chip source is measured to have a coincidence rate exceeding 1.5~MHz at a coincidence-to-accidental ratio (CAR) of 16.77 in the photon pair correlation experiment.
    The single-photon purity of this source, quantified by the heralded auto-correlation function $g^{(2)}_\text{h}(0)$, is measured by \rev{a} heralded Hanbury Brown--Twiss setup to reach the theoretical limit with the lowest value of $0.00094 \pm 0.00002$ obtained at a coincidence rate of 0.8~kHz and a CAR of 2220.
    The performance improvements are attributed to effective spectral filtering suppressing the noise as well as the coherent pump condition helped by optical injection locking.
    The reported results provide a reliable standard for benchmarking \rev{heralded single-photon sources} and present a state-of-the-art heralded source for quantum information processing.

\end{abstract}


\section{Introduction}

	Single photons are \rev{a} precious resource for quantum information applications~\cite{couteau_applications_2023}. 
	They can either be produced on-demand by quantum emitters~\cite{yu2023TelecombandQuantum} or heralded from spontaneous parametric photon-pair generation processes~\cite{kwiatUltrabrightSourcePolarizationentangled1999, chenTwophotonstateGenerationFourwave2005}. 
	The latter approach has the advantage of room-temperature operation and inherent compatibility with photonic waveguide integration~\cite{signoriniOnchipHeraldedSingle2020}.
    Particularly, achieving phase matching for spontaneous four-wave mixing (SFWM)---a third-order ($\chi^{(3)}$) nonlinear process---is relatively easy. 
	To date, a broad class of on-chip SFWM sources has been developed on various material platforms~\cite{doi:10.34133/adi.0032},
    among which silicon-on-insulator (SOI) attracts a great deal of interest due to its compatibility with mature complementary metal oxide semiconductor (CMOS) manufacturing technology.

    Single-photon purity is a key parameter for evaluating the performance of heralded single-photon sources (\rev{HSPS's}). 
    It is measurable by the \textit{heralded} auto-correlation function $g^{(2)}_\text{h}(0)$ using a heralded Hanbury Brown--Twiss (HBT) setup.
    While a value close to 0 is desirable for quantum applications, the practical $g^{(2)}_\text{h}(0)$ value of an HSPS \rev{has to trade-off} with the source brightness.
    Specifically, a stronger pump produces a \rev{higher rate} of heralded single photons, but at the expense of single-photon purity due to the increased probability of multi-pairs in the nonlinear process.
    To complicate further, single-photon purity is also strongly affected by the residual pump or spontaneous Raman photons leaked into the heralded path due to imperfect spectral filtering. 
    So far, despite a plethora of research achieved impressively low $g^{(2)}_\text{h}(0)$~\cite{du2024DemonstrationLow, maSiliconPhotonicEntangled2017,steinerUltrabrightEntangledPhotonPairGeneration2021}, there lacks a simple criterion to determine whether an HSPS has reached its theoretically allowed single-photon purity at a given brightness.

    To bridge this gap, starting from the photon number distribution, we derive an explicit limit for single-photon purity of \rev{an} HSPS under coherent pump condition.
    This limit is subsequently verified on a pulsed SFWM source, pumped by 2.5~GHz, 70~ps stable pulses that are generated by a pair of laser diodes in optical injection locking configuration~\cite{yuan2016Directly,zengControlledEntanglementSource2023}.  
    The nonlinear SFWM medium is a small-footprint spiral waveguide structure fabricated on SOI platform, \rev{and it} is designed to mitigate on-chip spontaneous Raman scattering.
    Our source is characterized to have brightness-limited single-photon purity for the pump power varying over two orders of magnitude.  
    Specifically, under a strong pump, we measure a $g_\text{h}^{(2)}(0)$ value of $0.086 \pm 0.002$ with a 0.8~MHz coincidence rate and a coincidence-to-accidental ratio (CAR) of 26.33.
    Given a weak pump, the source is measured to reach the lowest $g_\text{h}^{(2)}(0)$ value of $0.00094 \pm 0.00002$ together with a coincidence rate of 0.8~kHz and a CAR of 2220.
    Moreover, in terms of heralded photon brightness, we measured an unprecedentedly high coincidence rate of 1.51 MHz with a modest CAR of 16.77.

\section{Theoretical limit of $g^{(2)}_\text{h}(0)$}

    \textbf{Figure~\ref{g2CAR}(a)} schematically depicts a typical heralded HBT setup for measuring $g^{(2)}_\text{h}(0)$ of an HSPS.  
    A nonlinear medium, either $\chi ^{(2)}$ or $\chi ^{(3)}$ material, 
    is optically pumped to produce on average $\mu$ pairs of signal ($s$) and idler ($i$) photons per pump pulse when pulsed excitation is applied. 
    In the case of a continuous wave (CW) pump, $\mu$ denotes the average number of generated photon pairs per temporal bin adopted in the coincidence measurement.
    Signal and idler photons are \rev{separated spatially into different paths} via optical filtering components.
    Detecting a photon at one path heralds another photon in the other path as photons in each pair are simultaneously generated. 
    Single-photon purity of an HSPS is characterized by the heralded auto-correlation function~\cite{signoriniOnchipHeraldedSingle2020}
\begin{equation}\label{eq:trig2}
	g_\text{h}^{(2)}(0)=\frac{\left\langle\hat{a}_1^{\dagger} \hat{a}_2^{\dagger} \hat{a}_3^{\dagger} \hat{a}_3 \hat{a}_2 \hat{a}_1\right\rangle}{\left\langle\hat{a}_1^{\dagger} \hat{a}_2^{\dagger} 		\hat{a}_2 \hat{a}_1\right\rangle\left\langle\hat{a}_1^{\dagger} \hat{a}_3^{\dagger} \hat{a}_3 \hat{a}_1\right\rangle}\left\langle\hat{a}_1^{\dagger} \hat{a}_1\right\rangle,
\end{equation}
    where $a_j$ and $a_j^\dagger$ ($j=1,2,3$) are respectively photon annihilation and creation operators at port $j$.

    A coherent pump (CW or pulsed) produces Poissonian photon number distribution $P(n)=e^{-\mu}\frac{\mu^{n}}{n!}$ for signal and idler photons for a large number of distinguishable temporal modes~\cite{takesueEffectsMultiplePairs2010}.
    Note that the same results apply to other degrees of freedom such as spectral modes~\cite{christProbingMultimodeSqueezing2011}.
    From the above fact,  $g_\text{h}^{(2)}(0)$ can be derived to have an explicit dependence on the mean photon-pair number $\mu$ (see Methods)
\begin{equation}    \label{eq:possg2}
    g_\text{h}^{(2)}(0)=1-\frac{1}{(\mu+1)^2}.
\end{equation}
    With this equation, both $g^{(2)}_\text{h}(0)$ and $\mu$ are required to identify the performance gap of an HSPS from its theoretical limit. 

    We note that $g_\text{h}^{(2)}(0)$ is directly measurable.
    However, $\mu$ has to be inferred from other parameters including the channel collection efficiency, which is hard to measure precisely.
    Thus we need an alternate independent parameter that is directly measurable to construct the limit.
    As we are considering the theoretical limit, we assume that the noise photons are \rev{sufficiently} suppressed, which is technically feasible as we demonstrate in the subsequent section.
    Under this assumption, we can use CAR which quantifies the temporal coherence of generated signal and idler photons to estimate $\mu$ with high precision~\cite{takesueEffectsMultiplePairs2010}
\begin{equation} 	\label{eqCAR0}
	\text{CAR}	:=\frac{\text{coincidence counts}}{\text{accidental counts}}
    			=1 + \frac{1}{\mu}.
\end{equation}
    Experimentally, a coincidence count refers to an event of simultaneous detection (within a selected time window) by the single photon detectors (SPDs) connected to the signal and idler ports, whereas an accidental count refers to an uncorrelated event of the clicks, which manifests as the background counts in the coincidence measurement.
    Theoretically, CAR can also be written in terms of cross-correlation~\cite{signoriniOnchipHeraldedSingle2020}
\begin{equation}	\label{eqCAR1}
	\text{CAR}=
	\frac{\langle\hat{a}_{s}^{\dagger}\hat{a}_{i}^{\dagger}\hat{a}_{i}\hat{a}_{s}\rangle}
	{\langle\hat{a}_{s}^{\dagger}\hat{a}_{s}\rangle\langle\hat{a}_{i}^{\dagger}\hat{a}_{i}\rangle},
\end{equation}
	where $a_{s(i)}$ and $a_{s(i)}^\dagger$ are respectively annihilation and creation operators of signal and idler photons.
		The term $\langle\hat{a}_{s}^{\dagger}\hat{a}_{i}^{\dagger}\hat{a}_{i}\hat{a}_{s}\rangle$ represents the coincidence probability, which can be expressed as $\frac{C}{\epsilon_s\epsilon_i R}$, where $C$ is the raw coincidence rate, $\epsilon_{s(i)}$ is the collection efficiency of signal (idler) channel, and $R$ is pulse repetition rate (exemplified by a pulsed pump).
		While $\langle\hat{a}_{s}^{\dagger}\hat{a}_{s}\rangle$ and $\langle\hat{a}_{i}^{\dagger}\hat{a}_{i}\rangle$ is the photon click probabilities at the signal and idler ports, respectively, which equals to $\frac{S_{s(i)}}{\epsilon_{s(i)}R}$, where $S_{s(i)}$ is the raw single count rate.
        By a simple calculation, we have $\text{CAR}=\frac{CR}{S_{s}S_{i}}$ with the collection efficiency terms being canceled.
        
		Combining Equation~\eqref{eq:possg2} and \eqref{eqCAR0}, we obtain the theoretical limit for single-photon purity achievable with a coherent pump
\begin{equation}\label{eq:possg2car}
    g_\text{h}^{(2)}(0)=\frac{2\text{CAR}-1}{\text{CAR}^{2}}.
\end{equation}
    This limit applies to \rev{general parametric} HSPS's under a coherent pump, including both spontaneous parametric down-conversion (SPDC) and SWFM sources.
    
    Since both $g_\text{h}^{(2)}(0)$ and CAR are directly measurable and notably mutually compatible, Equation~\eqref{eq:possg2car} can be used to examine faithfully whether an HSPS has reached its brightness-limited single-photon purity.
    To showcase the usage of our bound, we compare in Figure~\ref{g2CAR}(b) the data (empty symbols) extracted from reported HSPS sources~\cite{maSiliconPhotonicEntangled2017,steinerUltrabrightEntangledPhotonPairGeneration2021,guoHighCoincidencetoaccidentalRatio2017,chunlexiongPhotonicCrystalWaveguide2015} against the theoretical limit (red line). 
    We identify that none of those sources has reached the limit, even when the CAR values exceeding $10^3$ were reported~\cite{footnote0}.
    Deviation from this limit can result from several factors, including detector dark counts, noise photons, and/or an incoherent pump.

    The impact of an incoherent pump can be modeled as follows.
    The incoherent condition results in intensity fluctuation, which can be described as the photon number distribution of the pump in a statistical mixture of Poissonian distributions with different values of mean photon number $\nu$~\cite{nakataIntensityFluctuationGainswitched2017}, 
\begin{equation}
    P(n)=\int d\nu f(\nu) e^{-\nu}\frac{\nu^{n}}{n!}. \nonumber
\end{equation}
    With this distribution, one can obtain $g_\text{h}^{(2)}(0)$ at arbitrary photon number distributions
\begin{equation}\label{g2h}
    g_\text{h}^{(2)}(0)=\frac{\langle\nu^3\rangle+2\langle\nu^2\rangle}{\left(\langle\nu^2\rangle+\langle\nu\rangle\right)^2}\langle\nu\rangle
    =
    \frac{g_\text{u}^{(3)}(0)\mu^2+2g_\text{u}^{(2)}(0)\mu}{(g_\text{u}^{(2)}(0)\mu+1)^2},
\end{equation}
    where $\mu=\langle\nu\rangle=\sum nP(n)$, while $g_\text{u}^{(2)}(0)$ and $g_\text{u}^{(3)}(0)$ \rev{are} the second\rev{-}order and third\rev{-}order correlation function\rev{s} of HSPS under \textit{unheralded} conditions, respectively.
    When $\mu\ll 1$, Equation~\eqref{g2h} is approximate\rev{d} to $g_\text{h}^{(2)}(0)=\frac{2g_\text{u}^{(2)}(0)\mu}{(g_\text{u}^{(2)}(0)\mu+1)^2}$, \rev{which allows us to} quantitatively evaluate the impact of intensity fluctuation on the single-photon purity of HSPS by additionally measuring $g_\text{u}^{(2)}(0)$.

\section{Experimental setup}

    We demonstrate that the theoretical limit is indeed attainable with a \rev{home}made HSPS.
    \textbf{Figure~\ref{experimental}(a)} shows our experimental setup. 
    A pair of semiconductor distributed feedback (DFB) laser diodes both emitting at a wavelength of 1550.52~nm are used in an optical injection locking configuration~\cite{yuan2016Directly}. 
    The master laser (ML) runs in CW mode. 
    Its emission is used to seed the slave laser (SL), which is gain-switched to produce 2.5~GHz, 70~ps laser pulses.
    Optical seeding reduces time jitter and improves intensity stability of the SL gain-switched pulses~\cite{comandar2016}.
    Switching off the ML allows us to purposely leave the SL gain-switched output in strong intensity fluctuation and thus demonstrate its effect on single-photon purity.
 
    The SL gain-switched pulses are amplified by an erbium-doped fiber amplifier (EDFA), followed by a fiber Bragg grating (FBG) with 0.4~nm full-width-at-half-maximum (FWHM) passband and a tunable bandpass filter (BPF) with 0.3~nm passband in order to remove the residual amplified spontaneous noise photons with $\geq$106~dB isolation. 
    The pump power is monitored by an optical power meter (PM) through a 10:90 tap beam splitter and then actively stabilized using a variable optical attenuator (VOA). 
    The pump pulses are aligned by a polarization controller (PC) to the TE mode of the silicon spiral waveguide.  
 
    In the spiral waveguide, two pump photons annihilate for the creation of two daughter photons of frequencies $\omega_{s}$ (signal) and $\omega_{i}$ (idler), which symmetrically surround the pump frequency $\omega_{p}$. 
    The signal/idler spectrum typically spans over 12~THz~\cite{haradaFrequencyPolarizationCharacteristics2010}, and we choose to use the International Telecommunication Union (ITU) channels C39 (1546.92~nm) and C28 (1554.94~nm) for the signal and idler outputs, respectively.
    To reject the residual pump photons, a dense wavelength divisional multiplexer (DWDM) which has 0.6~nm passband and non-adjacent channel isolation of no less than 105~dB, is applied after the spiral waveguide chip.
    The filtered photons are sent to a pair of superconducting nanowire single-photon detectors (SNSPDs) with their polarization controlled by a PC before each detector input for optimal detection efficiencies. 
    These SNSPDs feature about 71\% detection efficiency, $\sim$300~Hz dark count rate per channel, and 80~ps time jitter.
    Note that in measuring the heralded single-photon purity, one of the detectors is replaced by the HBT setup (dashed box).
    A multi-channel time-correlated single photon counting (TCSPC) instrument is used for correlation measurement.
    
    We present the design of our SOI chip in Figure~\ref{experimental}(b) and (c).
	Fabricated on a Si wafer, the spiral waveguide is designed for TE mode transmission and has a cross-section dimension of $450\times220$~{$\text{nm}^2$}. 
    In order to minimize transmission losses, each waveguide bend was designed to be no less than 22~$\mu$m. 
    Additionally, a fine annealing process during fabrication was performed to reduce the waveguide sidewall roughness after dry etching the Silicon layer \rev{during fabrication}.
    The total length of the waveguide is about 1.3 cm, with a transmission loss of 2.6~dB/cm. 
    Grating couplers are used on both input and output to couple light between fiber and waveguide with a loss of around 4.5~dB per facet.

\section{Results}

    We first characterize the auto-correlation function of the pump light using the HBT setup to verify whether it complies with Poissonian photon number statistics by bypassing the nonlinear material and attenuating the pump light to single-photon level.
    We start with the ML switched off and gain-switch the SL to output optical pulses at 2.5~GHz, which results in considerable intensity fluctuation.
    As revealed in \textbf{Figure~\ref{sf1}(a)}, the corresponding auto-correlation function is bunched at 0-delay, which is the typical behavior for gain-switched lasers at near-threshold operation~\cite{dynesTestingPhotonnumberStatistics2018}. 
    Noticeable coincidence dips at $\pm 0.4$~ns delays imply a negative correlation between adjacent laser pulses, as previously reported~\cite{nakataIntensityFluctuationGainswitched2017}.
    Turning on the ML and injecting $55$~$\mu$W light into SL, the SL output becomes stable in intensity as evidenced by the auto-correlation shown in Figure~\ref{sf1}(b). 
    We obtain $g_\text{p}^{(2)}(0) =  1.010 \pm 0.004$ in this case, approaching the expected value of $g_\text{p}^{(2)}(0) = 1$ for a coherent light source. 
    With a small deviation, we consider the pump source as coherent and obeys Poissonian photon number distribution.
    
    To evaluate the performance in terms of collected photon rates of the source, we perform the correlation measurement on the signal and idler photons.
    Under the coherent pump condition, we measure the single 
    count rate ($S$) and coincidence rate ($C$) as a function of the average pump power ($P$) entering the silicon spiral waveguide (i.e., the optical power after the fiber-to-chip coupling). 
    The associated heralding efficiency $\eta_{s(i)}$ can also be obtained with $\eta_{s(i)}=C/S_{i(s)}$.
    As shown in \textbf{Figure~\ref{SandC}(a)}, both the signal and idler single count rates show quadratic dependence on the pump power, which can be fit using~\cite{doi:10.34133/adi.0032} 
\begin{equation}
	S=aP^{2}+bP+c,\nonumber
\end{equation}  
    where $aP^{2}$, $bP$, and $c$ represent the count rates arising from SFWM photons, noise photons, and detector dark counts, respectively.  
    Coefficients $a$, $b$ and $c$ have the dimensions of MHz$\cdot$mW$^{-2}$, MHz$\cdot$mW$^{-1}$, and MHz, respectively, whose values are provided in the figure caption. 
    To avoid heralding efficiency saturation~\cite{Xiang:20} as indicated in Figure~\ref{SandC}(c), we limit the fitting region to less than 3.00~mW on-chip power, and the fitted results are shown as solid lines Figure~\ref{SandC}(a).
    According to the fitting coefficients, the noise photon, i.e., the linear term $bP$ dominates at relatively low pump powers, e.g., $P\leq 0.1$~mW.
    These noise photons may arise from the residual pump and Raman photons leaked into the signal or idler channel.
   	As the power increases, due to the quadratic behavior of the SFWM photons versus the linear behavior of noise photons, the SFWM photons gradually dominate the single counts.
 	At the saturation point---mostly caused by the two-photon absorption effect in silicon~\cite{linSiliconWaveguidesCreating2006}, we measure a maximum heralding efficiency of 11.3\% at the pump power of 5.68~mW \rev{for} channel C28.
    Note that the signal and idler channels do not have equal single count rates, which we attribute to the difference in the channel losses.
    \rev{Specifically, the respective losses of the DWDM channels C28 and C39 are measured to be 1.23 and 0.69 dB.}
    This inequality also leads to asymmetric heralding efficiencies between C28 and C39 channels as shown in Figure~\ref{SandC}(c).  

    Figure~\ref{SandC}(b) shows the measured coincidence rates, together with a polynomial fitting.   
    At a pump power of 7.10~mW, we obtain a coincidence rate of 1.51~MHz and a corresponding CAR of $16.77 \pm 0.01$.
    To our knowledge, this coincidence rate surpasses all previous SFWM results.
    Indeed, near-MHz coincidence rate was only achieved with an SPDC source but at a lower CAR of 7.5~\cite{Bock:16}. 
    In \textbf{Figure~\ref{fig11}}, we plot the measured CAR as a function of the coincidence rate denoted by hollow squares.

    \rev{To explain the observed monotonic decrease CAR with the coincidence rate, we resort to Equation (4) and use the heralding relation $S_{s(i)}= C/\eta_{i(s)}$ to derive the expression of CAR in terms of $C$ and $\eta_{s(i)}$ for a pulsed pump with a repetition rate $R$}
\begin{equation}\label{eqCAR}
	\text{CAR}=\frac{\eta_{s} \eta_{i} R}{C}.
\end{equation} 
    As expected from Equation~\eqref{eqCAR}, the CAR value decreases along with the increasing coincidence rate $C$, with the highest CAR being $4630 \pm 614$.
	Note that our data reproduces excellently the theoretical dependence of CAR on $C$ in Equation~\eqref{eqCAR} (the red solid curve).
	In the inset of Figure~\ref{fig11}, we present an example of a correlation histogram, which is measured at an on-chip power of 0.35~mW.
	In extracting the CAR value, we choose a time-bin width of $\tau=0.4~\text{ns}$ to cover the entire coincidence peak.
                   
\begin{figure}[h]
    \centering	
    \includegraphics[width=.5\columnwidth]{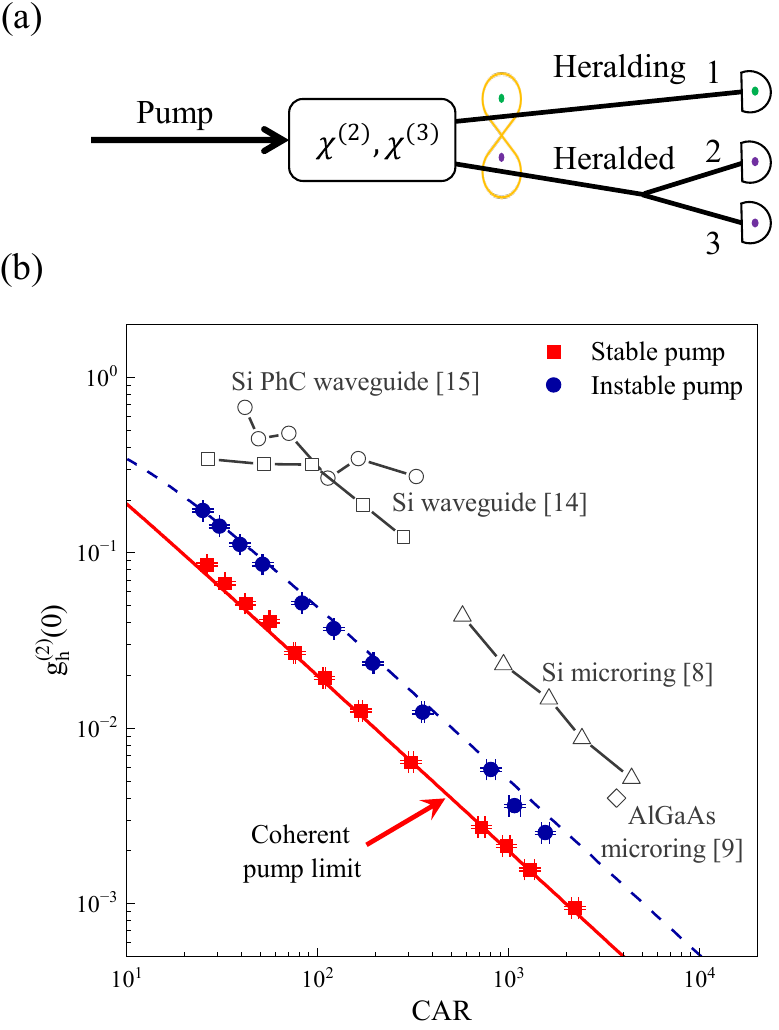}
    \caption{\label{g2CAR}
        (a) A schematic for measuring single-photon purity of HSPS.
        (b) $g_\text{h}^{(2)}(0)$ values, achieved in this work (solid symbols) or extracted from literature (open symbols). 
        The red solid line denotes the theoretical limit under coherent pump conditions, while the dashed line simulates an SFWM source under a pump with a certain degree of intensity fluctuation.
    }
\end{figure}

\begin{figure}
    \centering	
    \includegraphics[width=\columnwidth]{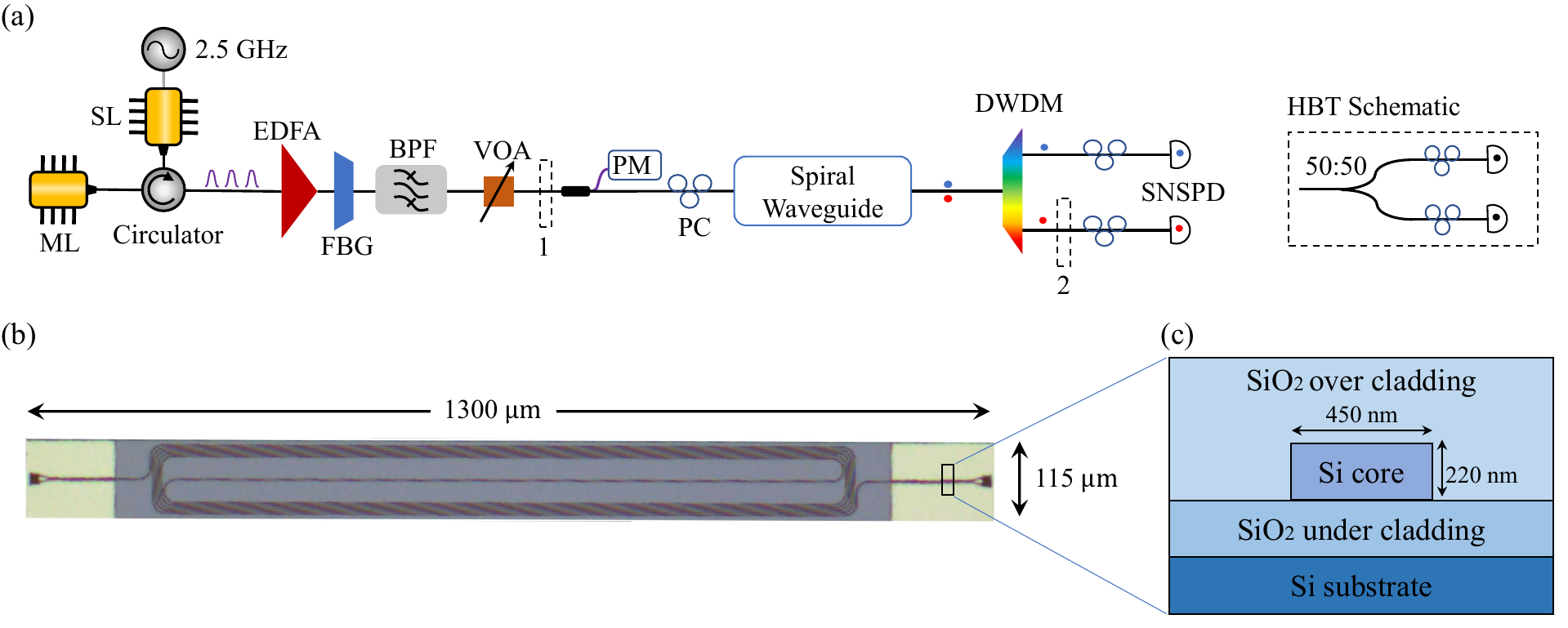}           
    \caption{\label{experimental}
        (a) Experimental setup.
        The HBT setup (dashed box) can be used at position 1 or 2.
        Position 1: to characterize the intensity fluctuation of the pump.
        Position 2: to measure the heralded single-photon purity of the channel.
        ML: master laser, SL: slave laser, EDFA: erbium-doped fiber amplifier, FBG: fiber Bragg grating, BPF: band-pass filter, VOA: variable optical attenuator, PM: power meter, PC: polarization controller, DWDM: dense wavelength division multiplexer, SNSPD: Superconducting nanowire single photon detector.
        (b) Image of the rectangular spiral waveguide by a microscope. 
        (c) Schematic of the waveguide cross-section.
    }	     
\end{figure}

\begin{figure}
	\centering	
	\includegraphics[width=.6\columnwidth]{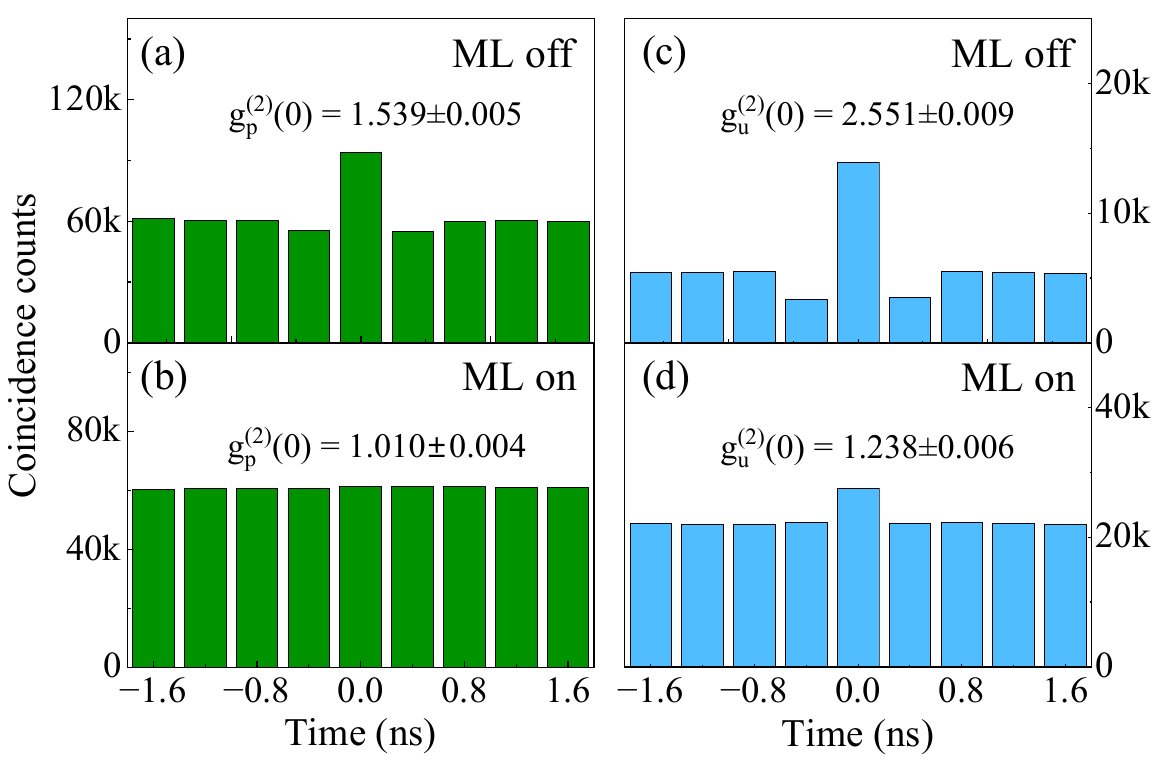}
	\caption{\label{sf1} 
		Auto-correlation function of the pump under (a) incoherent (master laser off) and (b) coherent (master laser on) conditions, 
		and the unheralded idler photons under (c) incoherent and (d) coherent conditions.
	}     
\end{figure}

\begin{figure} 
	\centering	
	\includegraphics[width=.6\columnwidth]{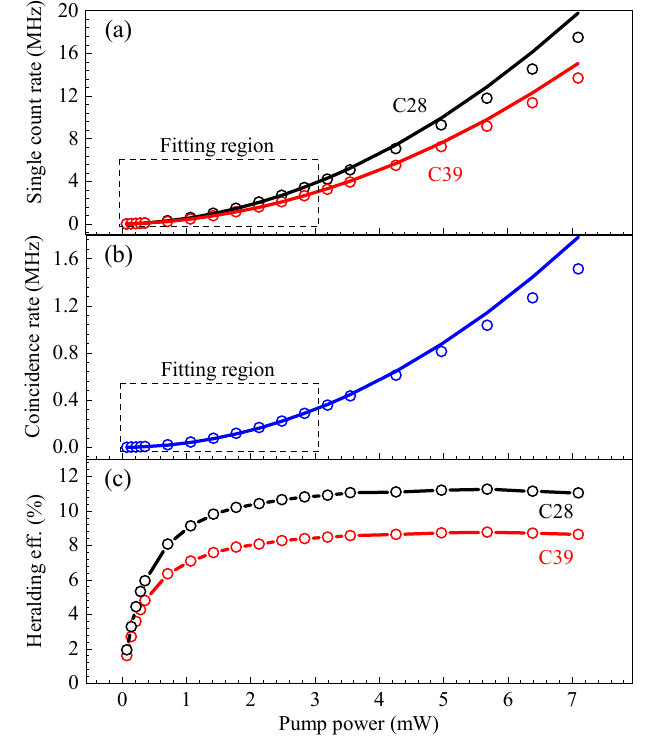}
	\caption{\label{SandC} 
		Experimental results (symbols) and polynomial fittings (lines) for
		(a) single count and (b) coincidence rates between DWDM channels C28 and C39. 
		The polynomial fitting formulas are
		$S_{s}=0.369P^2+0.171P+0.0003,~S_{i}=0.278P^2+0.150P+0.0003$ and $C=0.032P^2+0.002P$ for single count rate of signal (idler) photons and coincidence rate.
		(c) The measured heralding efficiency versus pump power.
	}     
\end{figure}

\begin{figure} 
\centering	
    \includegraphics[width=.6\columnwidth]{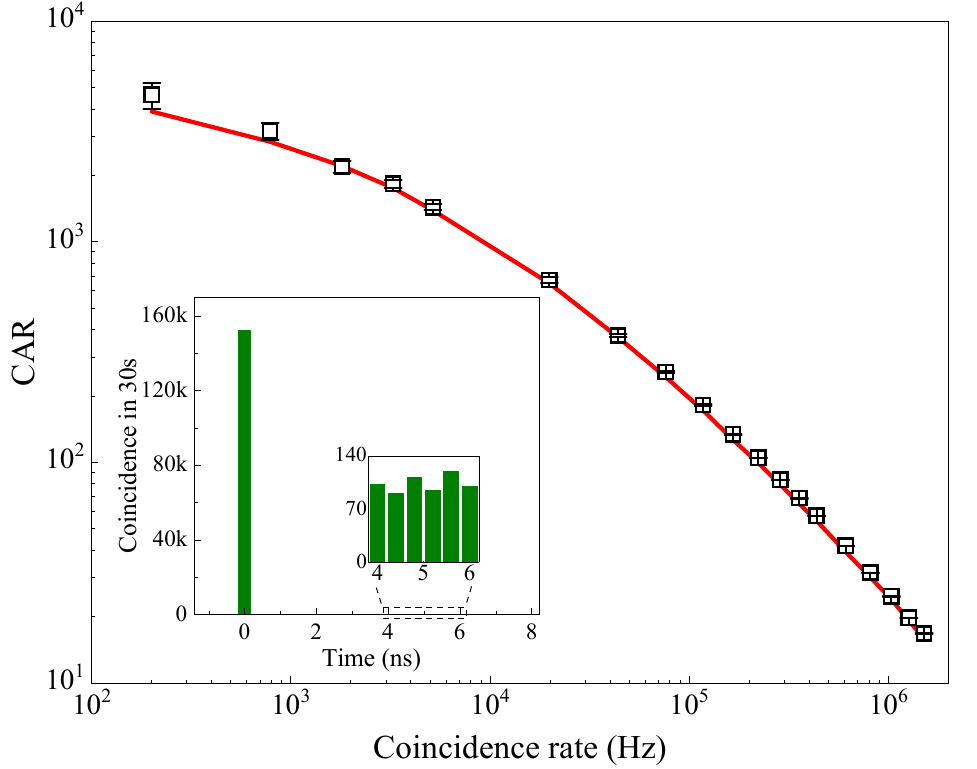}
    \caption{\label{fig11} 
        CAR versus coincidence rate: measured results (symbols) and simulation (line).
        Inset: An example of time-correlation histogram, measured at an on-chip power of 0.35~mW.
    }     
\end{figure}

    To compare against the theoretical limit for single-photon purity in Equation~\eqref{eq:possg2car}, we measure $g_\text{h}^{(2)}(0)$ under the coherent condition using the heralded HBT setup, which involves the three-fold coincidence measurement.
    Note that a CAR value can be extracted from the time stamp data of the heralding and one of the heralded detectors.
    The results together with error bars (red symbols) are plotted in Figure~\ref{g2CAR}(b).
    The lowest $g^{(2)}_\text{h}(0)$ \rev{is} measured as $0.00094 \pm 0.00002 $ with CAR equaling to $2220 \pm 90$, which is among the best $g^{(2)}_\text{h}(0)$ values reported to date for an HSPS.
    We note that an even lower $g^{(2)}_\text{h}(0)$ with a reasonably small uncertainty can \rev{in principle} be obtained with our source but at the cost of \rev{measurement} time, as this requires a lower pump power, which however is not the main pursuit of this work.
    For overall $g^{(2)}_\text{h}(0)$ versus CAR relation, the experimental data indeed reach the theoretical limit, suggesting that our source produces the ideal single-photon purity permitted by a \rev{parametric heralded} source.
    We attribute this achievement to both effective spectral filtering~\cite{footnote1} and the coherent pump condition helped by optical injection locking.
    
    As we have mentioned \rev{earlier}, deviation from coherent pump conditions can deteriorate the single-photon purity for a given CAR value.
    To demonstrate, we switch off the ML to introduce intensity fluctuation, leading to the pump to have a $g_\text{p}^{(2)}(0) = 1.539 \pm 0.005$ which corresponds to the condition shown in Figure~\ref{sf1}(a).
    Then, a stronger bunching of $g_\text{u}^{(2)}(0) = 2.551\pm 0.009$ is measured in the unheralded idler channel.
    It is reasonable that the fluctuation in $g_\text{p}^{(2)}(0)$ can results in an amplified fluctuation in $g_\text{u}^{(2)}(0)$, as SFWM is a third-order nonlinear process.
    It is worth mentioning that the measured $g_\text{u}^{(2)}(0)$ can go beyond the thermal-like statistics limit~\cite{takesueEffectsMultiplePairs2010} as the pump is no longer coherent.
    As a comparison, we measure a $g_\text{u}^{(2)}(0)=1.238\pm0.006$ under coherent condition as shown in Figure~\ref{sf1}(d), which is consistent with our analysis.
    Due to the instable pump, the heralded $g^{(2)}_\text{h}(0)$ values become considerably worse compared to the coherent condition, which is exhibited in Figure~\ref{g2CAR}(b) denoted by blue dots.  
    The measured data are in good agreement with the theoretical prediction (dashed line) taking the unheralded $g_\text{u}^{(2)}(0)=2.551$ in Equation~\eqref{g2h}.

    As summarized in Figure~\ref{g2CAR}(b), previous HSPS sources have $g^{(2)}_\text{h}(0)$ values that are worse than the theoretical limit even with impressive $g^{(2)}_\text{h}(0)$ and CAR values were obtained. 
    To further appreciate the advance of our source, we list the state-of-the-art SFWM photon pair sources under pulsed excitation in Table~\ref{tab:table1}.
        



\section{Conclusion}

    We have derived a theoretical limit for single-photon purity of general parametric HSPS's under coherent pump conditions, and subsequently verified it on a \rev{home}made on-chip SFWM source based on SOI platform.
    Our source has also achieved advances in \rev{individual} figures of merit.
    The source exhibits an unprecedented coincidence rate of 1.51~MHz together with a $\text{CAR}$ of 16.77.
    Moreover, we measure an ultra low $g^{(2)}_\text{h}(0)$ value of $0.00094 \pm 0.00002$.
    Thanks to its pulsed operation, we expect our high-quality HSPS to be useful in quantum information applications.
      

\section{Methods}
\subsection*{Derivation of the theoretical second-order correlation function related to $\mu$} 
	For a heralded single photon source (HSPS), the heralded second-order correlation function is
\begin{equation}
 	g_\text{h}^{(2)}(0)=\frac{\left\langle\hat{a}_1^{\dagger} \hat{a}_2^{\dagger} \hat{a}_3^{\dagger} \hat{a}_3 \hat{a}_2 \hat{a}_1\right\rangle}{\left\langle\hat{a}_1^{\dagger} \hat{a}_2^{\dagger} \hat{a}_2 \hat{a}_1\right\rangle\left\langle\hat{a}_1^{\dagger} \hat{a}_3^{\dagger} \hat{a}_3 \hat{a}_1\right\rangle}\left\langle\hat{a}_1^{\dagger} \hat{a}_1\right\rangle,
\end{equation}
    where the subscriptions denote the detectors at the heralding (1) or heralded (2,3) side.
   	As measurements between the heralding side and the heralded are independent, operators $\hat{a}_{1}$ and $\hat{a}_{2,3}$ are mutually commutative. 
	Therefore, $g_\text{h}^{(2)}(0)$ can be rewritten as
\begin{equation}\label{gh2}
	g_\text{h}^{(2)}(0)=\frac{\langle\hat{n_{s}} \hat{n_{i}}(\hat{n_{i}}-1) \rangle}{\langle\hat{n_{s}}\hat{n_{i}}\rangle^{2}}{\langle\hat{n_{s}}\rangle}, 
\end{equation}
	where $\hat{n}$ represents the photon number operator, the angle brackets represent the expected value, and the subscriptions $s$ and $i$ refer to the signal photons used for heralding and the idler photons being heralded, respectively.

	Due to the simultaneity of photon pairs generated through SFWM, the photon pair state can be written as~\cite{signoriniOnchipHeraldedSingle2020}
\begin{equation}\label{phi}
	|\Phi\rangle=\sum_{n=0}^{\infty}{P(n)|n,n\rangle_{s,i}}, 
\end{equation}
	where $P(n)$ is the probability of generating n photon pairs per pulse.

	Then, we can calculate the second-order correlation function of the heralded single-photon source $g_\text{h}^{(2)}(0)$ under any distribution by taking Equation~\eqref{phi} into Equation~\eqref{gh2}, and we obtain
\begin{equation}
	\begin{aligned}
	g_\text{h}^{(2)}(0)&=\frac{{\langle\Phi|\hat{n_{s}} \hat{n_{i}}(\hat{n_{i}}-1)|\Phi\rangle}}{\langle\Phi|\hat{n_{s}}\hat{n_{i}}|\Phi\rangle^{2}}{\langle\Phi|\hat{n_{s}}|\Phi\rangle}\\
	&=\frac{\sum{n^2(n-1)P(n)}}{(\sum {n^2P(n)})^2}\sum{nP(n)}.  \nonumber
	\end{aligned}
\end{equation}  
	Therefore, for SFWM sources with Poissonian distribution, $P(n)=e^{-\mu}\frac{\mu^{n}}{n!}$, its $g_\text{h}^{(2)}(0)$ can be expressed as
\begin{equation}
	g_\text{h}^{(2)}(0)=\frac{\mu^2(\mu+2)}{[\mu(\mu+1)]^2}\mu=1-\frac{1}{(\mu+1)^2}.  \nonumber
\end{equation}



\bigskip
\textbf{Supporting Information} \par 
N/A

\bigskip
\textbf{Acknowledgements} \par 
H. Wang and H. Yuan contributed equally to this work. 
This work was supported by National Natural Science Foundation of China under Grants 12105010 (Q. Z.), 62105034 (L. Z.), and 62250710162 (Z. Y.).

\bigskip
\textbf{Conflict of Interest}\par
	The authors declare no conflict of interest.

\bigskip
\textbf{Data Availability Statement}\par
	The data supporting the plots within this paper and other study findings are available from corresponding authors upon reasonable request.  

\bigskip
\textbf{Author Contributions}	\par
    Q.Z. and H.Y.W. devised the experimental setup. H.Y.W. and Q.Z. performed the experiment. H.H.Y. and Q.Z. designed, and H.H.Y. fabricated the silicon chip. All authors contributed to the discussion and writing of the manuscript. Z.Y. guided the project.

\bigskip

%
\bibliographystyle{MSP}

\begin{thebibliography}{10}
\providecommand{\url}[1]{\texttt{#1}}
\providecommand{\urlprefix}{URL }

\bibitem{couteau_applications_2023}
C.~Couteau, S.~Barz, T.~Durt, T.~Gerrits, J.~Huwer, R.~Prevedel, J.~Rarity, A.~Shields, G.~Weihs,
\newblock \emph{Nat. Rev. Phys.} \textbf{2023}, \emph{5} 326.

\bibitem{yu2023TelecombandQuantum}
Y.~Yu, S.~Liu, C.-M. Lee, P.~Michler, S.~Reitzenstein, K.~Srinivasan, E.~Waks, J.~Liu,
\newblock \emph{Nat. Nanotechnol.} \textbf{2023}, \emph{18} 1389.

\bibitem{kwiatUltrabrightSourcePolarizationentangled1999}
P.~G. Kwiat, E.~Waks, A.~G. White, I.~Appelbaum, P.~H. Eberhard,
\newblock \emph{Phys. Rev. A} \textbf{1999}, \emph{60} R773.

\bibitem{chenTwophotonstateGenerationFourwave2005}
J.~Chen, X.~Li, P.~Kumar,
\newblock \emph{Phys. Rev. A} \textbf{2005}, \emph{72} 033801.

\bibitem{signoriniOnchipHeraldedSingle2020}
S.~Signorini, L.~Pavesi,
\newblock \emph{AVS Quantum Sci.} \textbf{2020}, \emph{2} 041701.

\bibitem{doi:10.34133/adi.0032}
H.~Wang, Q.~Zeng, H.~Ma, Z.~Yuan,
\newblock \emph{Adv. Dev. Instrum.} \textbf{2024}, \emph{5} 0032.

\bibitem{du2024DemonstrationLow}
J.~Du, G.~F.~R. Chen, H.~Gao, J.~A. Grieve, D.~T.~H. Tan, A.~Ling,
\newblock \emph{Opt. Express} \textbf{2024}, \emph{32} 11406.

\bibitem{maSiliconPhotonicEntangled2017}
C.~Ma, X.~Wang, V.~Anant, A.~D. Beyer, M.~D. Shaw, S.~Mookherjea,
\newblock \emph{Opt. Express} \textbf{2017}, \emph{25} 32995.

\bibitem{steinerUltrabrightEntangledPhotonPairGeneration2021}
T.~J. Steiner, J.~E. Castro, L.~Chang, Q.~Dang, W.~Xie, J.~Norman, J.~E. Bowers, G.~Moody,
\newblock \emph{PRX Quantum} \textbf{2021}, \emph{2} 010337.

\bibitem{yuan2016Directly}
Z.~L. Yuan, B.~Fr{\"o}hlich, M.~Lucamarini, G.~L. Roberts, J.~F. Dynes, A.~J. Shields,
\newblock \emph{Phys. Rev. X} \textbf{2016}, \emph{6} 031044.

\bibitem{zengControlledEntanglementSource2023}
Q.~Zeng, H.~Wang, H.~Yuan, Y.~Fan, L.~Zhou, Y.~Gao, H.~Ma, Z.~Yuan,
\newblock \emph{Phys. Rev. Appl.} \textbf{2023}, \emph{19} 054048.

\bibitem{takesueEffectsMultiplePairs2010}
H.~Takesue, K.~Shimizu,
\newblock \emph{Opt. Commun.} \textbf{2010}, \emph{283} 276.

\bibitem{christProbingMultimodeSqueezing2011}
A.~Christ, K.~Laiho, A.~Eckstein, K.~N. Cassemiro, C.~Silberhorn,
\newblock \emph{New J. Phys.} \textbf{2011}, \emph{13} 033027.

\bibitem{guoHighCoincidencetoaccidentalRatio2017}
K.~Guo, E.~N. Christensen, J.~B. Christensen, J.~G. Koefoed, D.~Bacco, Y.~Ding, H.~Ou, K.~Rottwitt,
\newblock \emph{Appl. Phys. Express} \textbf{2017}, \emph{10} 062801.

\bibitem{chunlexiongPhotonicCrystalWaveguide2015}
{C. Xiong}, M.~J. Collins, M.~J. Steel, T.~F. Krauss, B.~J. Eggleton, A.~S. Clark,
\newblock \emph{IEEE J. Sel. Top. Quantum Electron.} \textbf{2015}, \emph{21} 205.

\bibitem{footnote0}
Reference [8] reported a record-high CAR up to $10^4$ but did not provide the corresponding single-photon purity.

\bibitem{nakataIntensityFluctuationGainswitched2017}
K.~Nakata, A.~Tomita, M.~Fujiwara, K.-i. Yoshino, A.~Tajima, A.~Okamoto, K.~Ogawa,
\newblock \emph{Opt. Express} \textbf{2017}, \emph{25} 622.

\bibitem{comandar2016}
L.~C. Comandar, M.~Lucamarini, B.~Fr\"ohlich, J.~F. Dynes, Z.~L. Yuan, A.~J. Shields,
\newblock \emph{Opt. Express} \textbf{2016}, \emph{24} 17849 .

\bibitem{haradaFrequencyPolarizationCharacteristics2010}
K.-I. Harada, H.~Takesue, H.~Fukuda, T.~Tsuchizawa, T.~Watanabe, K.~Yamada, Y.~Tokura, S.-I. Itabashi,
\newblock \emph{IEEE J. Sel. Top. Quantum Electron.} \textbf{2010}, \emph{16} 325.

\bibitem{dynesTestingPhotonnumberStatistics2018}
J.~F. Dynes, M.~Lucamarini, K.~A. Patel, A.~W. Sharpe, M.~B. Ward, Z.~L. Yuan, A.~J. Shields,
\newblock \emph{Opt. Express} \textbf{2018}, \emph{26} 22733.

\bibitem{Xiang:20}
C.~Xiang, W.~Jin, J.~Guo, C.~Williams, A.~M. Netherton, L.~Chang, P.~A. Morton, J.~E. Bowers,
\newblock \emph{Opt. Express} \textbf{2020}, \emph{28} 19926.

\bibitem{linSiliconWaveguidesCreating2006}
Q.~Lin, G.~P. Agrawal,
\newblock \emph{Opt. Lett.} \textbf{2006}, \emph{31} 3140.

\bibitem{Bock:16}
M.~Bock, A.~Lenhard, C.~Chunnilall, C.~Becher,
\newblock \emph{Opt. Express} \textbf{2016}, \emph{24} 23992.

\bibitem{footnote1}
In our setup, we chose a wide passband (0.6 nm) spectral filter that can effectively suppress the residual pump noise while incurring a relatively low insertion loss. In particular, we avoid the use of an ultra-narrow band filter because an overly strict filtering could impair the single-photon purity due to the incompatibility of frequency and time measurements.

\bibitem{WOS:000886923500005}
J.~Monteleone, III, M.~van Niekerk, M.~Ciminelli, G.~Leake, D.~Coleman, M.~Fanto, S.~Preble, In C.~Soci, M.~Sheldon, M.~Agio, I.~Aharonovich, editors,
Conference on Quantum Nanophotonic Materials,
\newblock Devices, and Systems Part of SPIE Nanoscience and Engineering Conference,
\emph{Proceedings of SPIE}. SPIE,
\textbf{2022}, San Diego, CA.

\bibitem{zhangCorrelatedPhotonPair2016}
X.~Zhang, Y.~Zhang, C.~Xiong, B.~J. Eggleton,
\newblock \emph{J. Opt.} \textbf{2016}, \emph{18} 074016.

\bibitem{choiCorrelatedPhotonPair2020}
J.~W. Choi, B.-U. Sohn, G.~F. Chen, D.~K. Ng, D.~T. Tan,
\newblock \emph{Opt. Commun.} \textbf{2020}, \emph{463} 125351.

\bibitem{clarkHeraldedSinglephotonSource2013}
A.~S. Clark, C.~Husko, M.~J. Collins, G.~Lehoucq, S.~Xavier, A.~De~Rossi, S.~Combri{\'e}, C.~Xiong, B.~J. Eggleton,
\newblock \emph{Opt. Lett.} \textbf{2013}, \emph{38} 649.

\end{thebibliography}




\begin{table*}[b]
    \caption{\label{tab:table1} State-of-the-art SFWM photon pair sources under pulsed/CW excitation.}
    
    \begin{tabular}[htbp]{ccccccc}

    \hline
    
    Ref (year) & Material & Repetition rate & Max CAR  &Min $g^{(2)}_\text{h}(0)~(C)$& Max raw $C$ (CAR)\\ 

    \hline
    
    this work&Si waveguide&2.5 GHz & 4630 
    & $0.00094 \pm 0.00002$ (0.80 kHz)  & 1.5 MHz (16.77)    \\

    ref.~\cite{WOS:000886923500005} (2022)&Si waveguide&500 MHz & 8 
    & -    & 300 kHz (8)         \\


    ref.~\cite{chunlexiongPhotonicCrystalWaveguide2015} (2015)& Si PhC waveguide & 50 MHz &  329 
    &      $<0.1$ (10 Hz)    & 48 Hz (41)     \\

    ref.~\cite{zhangCorrelatedPhotonPair2016} (2016)&$\text{Si}_{3}\text{N}_{4}$ waveguide & 50 Mhz & 16 & - & 1.6 kHz (10.2) \\

    ref.~\cite{choiCorrelatedPhotonPair2020} (2020)&USRN waveguide & 20 MHz & 7 &- & 1.8 Hz (2.6) \\

    ref.~\cite{clarkHeraldedSinglephotonSource2013} (2013)&GaInP PhC waveguide&50 MHz&63&0.06 (-)&4.5 Hz (12)\\

    ref.~\cite{guoHighCoincidencetoaccidentalRatio2017} (2017)& Si waveguide & CW & 673 & $0.12 \pm 0.09$ (2.68 kHz) & 22 kHz (27)\\

    ref.~\cite{maSiliconPhotonicEntangled2017} (2017) & Si ring  &    CW     & 12105  &  $0.00533 \pm 0.021$ (0.35 kHz) &  5 kHz (573)   \\

    ref.~\cite{steinerUltrabrightEntangledPhotonPairGeneration2021} (2021)& AlGaAs ring& CW & 4389 & $0.004 \pm 0.01$ (0.11 Hz)&2.0 Hz (354)\\


    \hline
    
  \end{tabular}
\end{table*}

\end{document}